\documentclass[runningheads]{llncs}
\usepackage{graphicx}
\usepackage{amssymb,amsfonts}
\usepackage{hyperref}
\usepackage{multirow}
\usepackage{color}
\usepackage{marvosym}
\begin{document}
\title{EGE-UNet: an Efficient Group Enhanced UNet for skin lesion segmentation}
%
%
\author{Jiacheng Ruan 
\and
Mingye Xie 
\and
Jingsheng Gao 
\and
Ting Liu 
\and
Yuzhuo Fu 
\textsuperscript{(\Letter)}\thanks{Yuzhuo Fu is the corresponding author and this work was partially supported by the National Natural Science Foundation of China (Grant No. 61977045).}}
\authorrunning{J. Ruan et al.}
%
\institute{Shanghai Jiao Tong University, China \\
\email{\{jackchenruan, xiemingye, gaojingsheng, louisa\_liu, yzfu\}@sjtu.edu.cn}}
\maketitle              
\begin{abstract}
Transformer and its variants have been widely used for medical image segmentation. However, the large number of parameter and computational load of these models make them unsuitable for mobile health applications. To address this issue, we propose a more efficient approach, the Efficient Group Enhanced UNet (\textbf{EGE-UNet}). We incorporate a Group multi-axis Hadamard Product Attention module (GHPA) and a Group Aggregation Bridge module (GAB) in a lightweight manner. The GHPA groups input features and performs Hadamard Product Attention mechanism (HPA) on different axes to extract pathological information from diverse perspectives. The GAB effectively fuses multi-scale information by grouping low-level features, high-level features, and a mask generated by the decoder at each stage. Comprehensive experiments on the ISIC2017 and ISIC2018 datasets demonstrate that EGE-UNet outperforms existing state-of-the-art methods. In short, compared to the TransFuse, our model achieves superior segmentation performance while reducing parameter and computation costs by \textbf{494x} and \textbf{160x}, respectively. Moreover, to our best knowledge, this is the first model with a parameter count limited to just \textbf{50KB}. Our code is available at \href{https://github.com/JCruan519/EGE-UNet}{https://github.com/JCruan519/EGE-UNet}.

\keywords{Medical image segmentation \and Light-weight model \and mobile health.}
\end{abstract}
\section{Introduction}

\begin{figure}[!t]
	\centerline{\includegraphics[width=13cm]{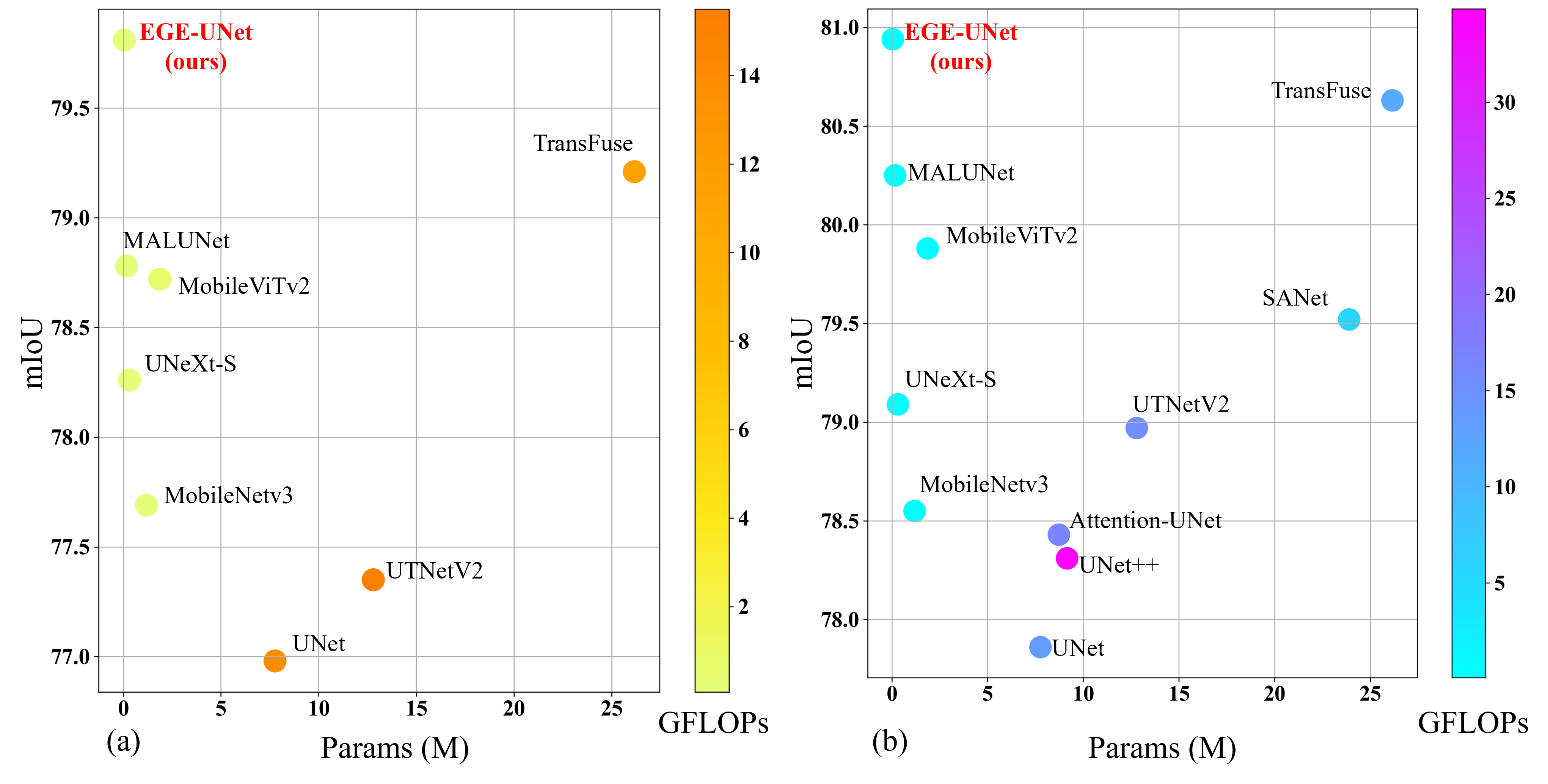}}
	\caption{(a) and (b) respectively show the visualization of comparative experimental results on the ISIC2017 and ISIC2018 datasets. The X-axis represents the number of parameters (lower is better), while Y-axis represents mIoU (higher is better). The color depth represents computational complexity (GFLOPs, lighter is better).} \label{fig1}
\end{figure}

Malignant melanoma is one of the most rapidly growing cancers in the world. As estimated by the American Cancer Society, there were approximately 100,350 new cases and over 6,500 deaths in 2020 \cite{skincancer}. Thus, an automated skin lesion segmentation system is imperative, as it can assist medical professionals in swiftly identifying lesion areas and facilitating subsequent treatment processes. To enhance the segmentation performance, recent studies tend to employ modules with larger parameter and computational complexity, such as incorporating self-attention mechanisms of Vision Transformer (ViT) \cite{vit}. For example, Swin-UNet \cite{swinunet}, based on the Swin Transformer \cite{swintrm}, leverages the feature extraction ability of self-attention mechanisms to improve segmentation performance. TransUNet \cite{transunet} has pioneered a serial fusion of CNN and ViT for medical image segmentation. TransFuse \cite{transfuse} employs a dual-path structure, utilizing CNN and ViT to capture local and global information, respectively. UTNetV2 \cite{utnetv2} utilizes a hybrid hierarchical architecture, efficient bidirectional attention, and semantic maps to achieve global multi-scale feature fusion, combining the strengths of CNN and ViT. TransBTS \cite{transbts} introduces self-attention into brain tumor segmentation tasks and uses it to aggregate high-level information.

Prior works have enhanced performance by introducing intricate modules, but neglected the constraint of computational resources in real medical settings. Hence, there is an urgent need to design a low-parameter and low-computational load model for segmentation tasks in mobile healthcare. Recently, UNeXt \cite{unext} has combined UNet \cite{unet} and MLP \cite{mlp} to develop a lightweight model that attains superior performance, while diminishing parameter and computation. Furthermore, MALUNet \cite{malunet} has reduced the model size by declining the number of model channels and introducing multiple attention modules, resulting in better performance for skin lesion segmentation than UNeXt. However, while MALUNet greatly reduces the number of parameter and computation, its segmentation performance is still lower than some large models, such as TransFuse. Therefore, in this study, we propose EGE-UNet, a lightweight skin lesion segmentation model that achieves state-of-the-art while significantly reducing parameter and computation costs. Additionally, to our best knowledge, this is the first work to reduce parameter to approximately \textbf{50KB}.

To be specific, EGE-UNet leverages two key modules: the Group multi-axis Hadamard Product Attention module (GHPA) and Group Aggregation Bridge module (GAB). On the one hand, recent models based on ViT \cite{vit} have shown promise, owing to the multi-head self-attention mechanism (MHSA). MHSA divides the input into multiple heads and calculates self-attention in each head, which allows the model to obtain information from diverse perspectives, integrate different knowledge, and improve performance. Nonetheless, the quadratic complexity of MHSA enormously increases the model's size. Therefore, we present the Hadamard Product Attention mechanism (HPA) with linear complexity. HPA employs a learnable weight and performs a hadamard product operation with the input to obtain the output. Subsequently, inspired by the multi-head mode in MHSA, we propose GHPA, which divides the input into different groups and performs HPA in each group. However, it is worth noting that we perform HPA on different axes in different groups, which helps to further obtain information from diverse perspectives. On the other hand, for GAB, since the size and shape of segmentation targets in medical images are inconsistent, it is essential to obtain multi-scale information \cite{malunet}. Therefore, GAB integrates high-level and low-level features with different sizes based on group aggregation, and additionally introduce mask information to assist feature fusion. Via combining the above two modules with UNet, we propose EGE-UNet, which achieves excellent segmentation performance with extremely low parameter and computation. Unlike previous approaches that focus solely on improving performance, our model also prioritizes usability in real-world environments. A clear comparison of EGE-UNet with others is shown in Figure \ref{fig1}.

\begin{figure}[!t]
	\centerline{\includegraphics[width=12.5cm]{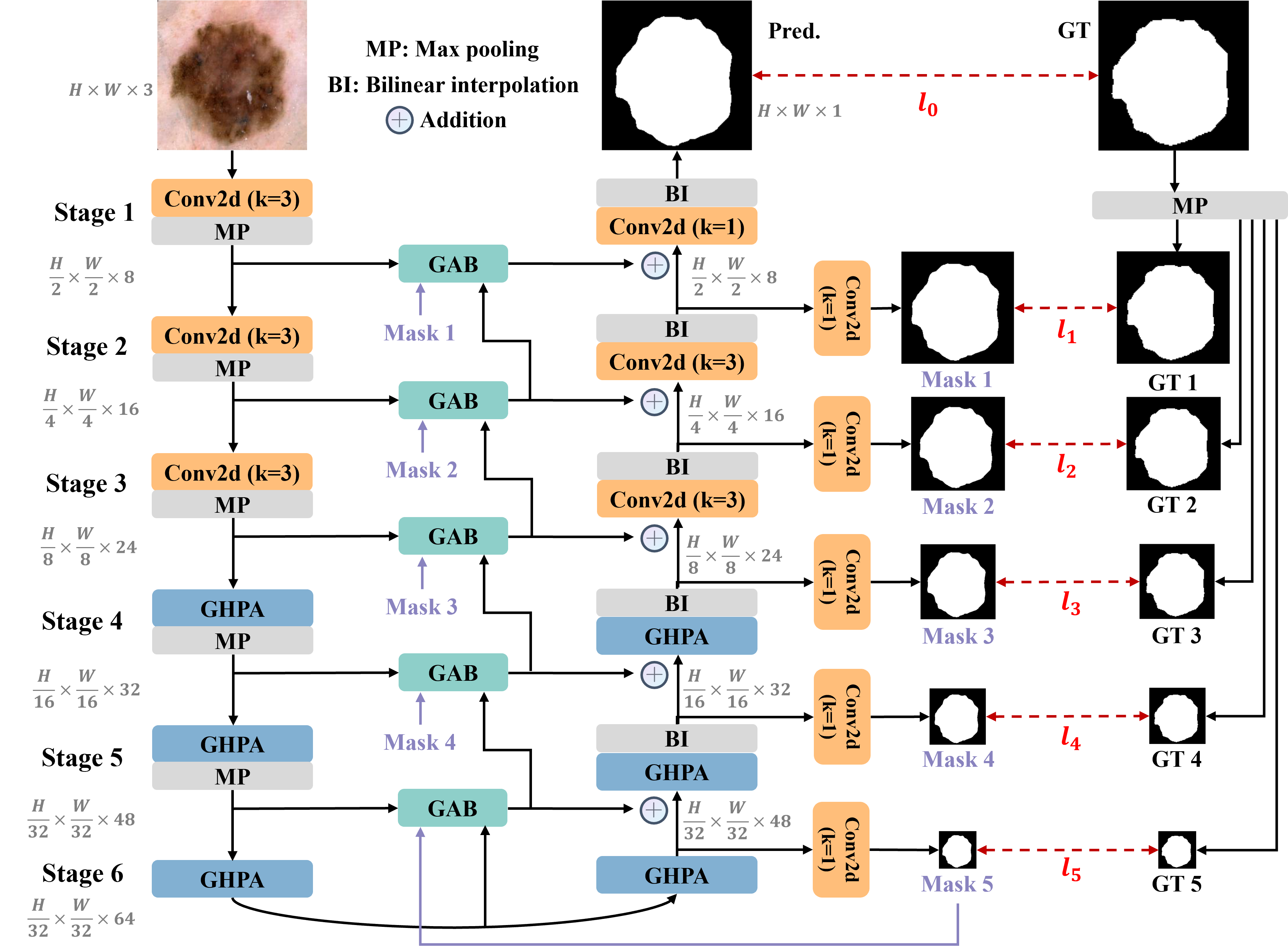}}
	\caption{The overview of EGE-UNet.} \label{fig2}
\end{figure}

In summary, our contributions are threefold: (1) GHPA and GAB are proposed, with the former efficiently acquiring and integrating multi-perspective information and the latter accepting features at different scales, along with an auxiliary mask for efficient multi-scale feature fusion. (2) We propose EGE-UNet, an extremely lightweight model designed for skin lesion segmentation. (3) We conduct extensive experiments, which demonstrate the effectiveness of our methods in achieving state-of-the-art performance with significantly lower resource requirements.

\begin{figure}[!t]
	\centerline{\includegraphics[width=9cm]{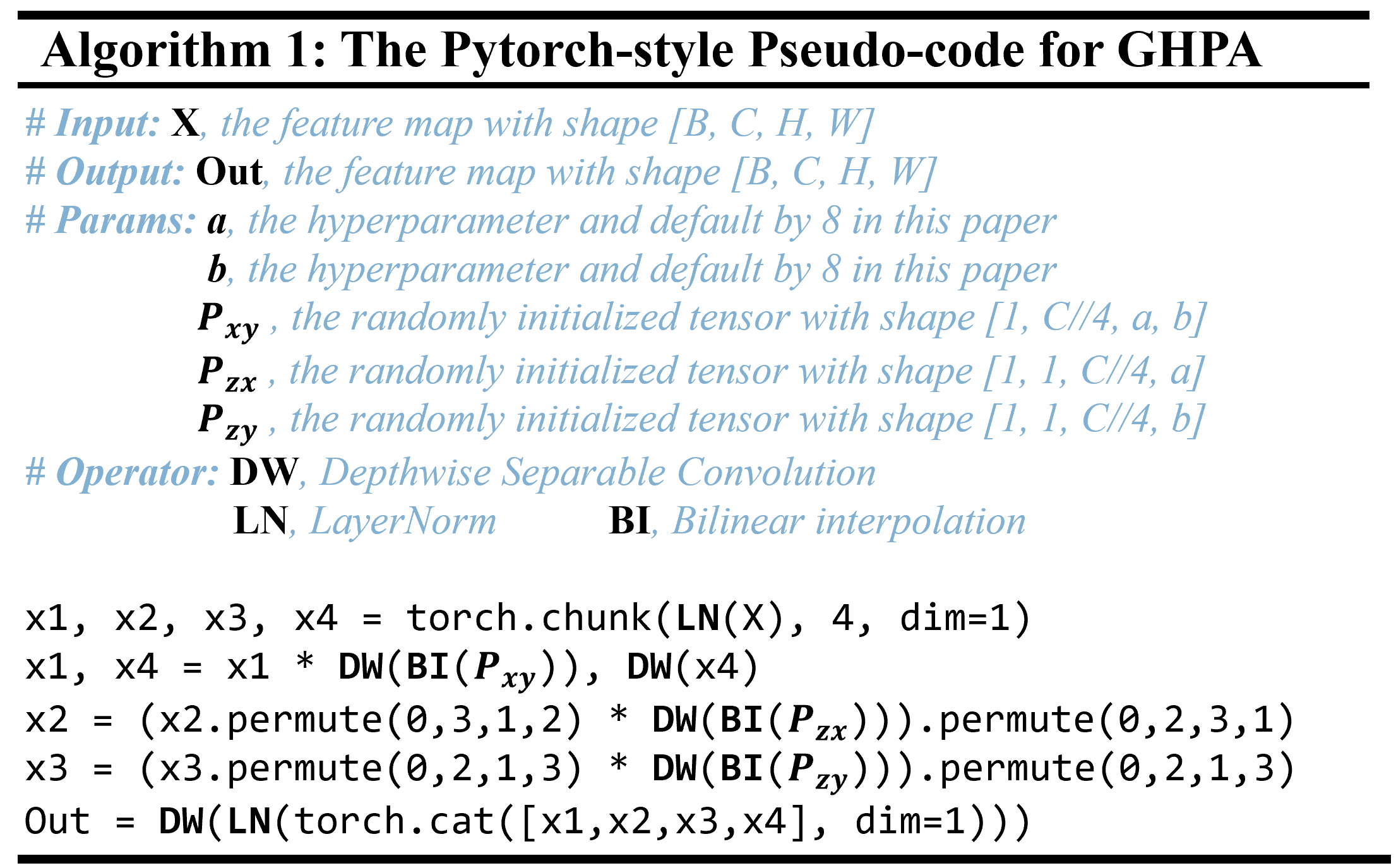}}
	\label{alg1}
\end{figure}

\section{EGE-UNet}

\subsubsection{The overall architecture.} EGE-UNet is illustrated in Figure \ref{fig2}, which is built upon the U-Shape architecture consisting of symmetric encoder-decoder parts. We take encoder part as an example. The encoder is composed of six stages, each with channel numbers of \{8, 16, 24, 32, 48, 64\}. While the first three stages employ plain convolutions with a kernel size of 3, the last three stages utilize the proposed GHPA to extract representation information from diverse perspectives. In contrast to the simple skip connections in UNet, EGE-UNet incorporates GAB for each stage between the encoder and decoder. Furthermore, our model leverages deep supervision \cite{unet++} to generate mask predictions of varying scales, which are utilized for loss function and serve as one of the inputs to GAB. Via the integration of these advanced modules, EGE-UNet significantly reduces the parameter and computational load while enhancing the segmentation performance compared to prior approaches.

\subsubsection{Group multi-axis Hadamard Product Attention module.}

To overcome the quadratic complexity issue posed by MHSA, we propose HPA with linear complexity. Given an input $x$ and a randomly initialized learnable tensor $p$, bilinear interpolation is first utilized to resize $p$ to match the size of $x$. Then, we employ depth-wise separable convolution (DW) \cite{mobilenets}\cite{mobilenetv2} on $p$, followed by a hadamard product operation between $x$ and $p$ to obtain the output. However, utilizing simple HPA alone is insufficient to extract information from multiple perspectives, resulting in unsatisfactory results. Motivated by the multi-head mode in MHSA, we introduce GHPA based on HPA, as illustrated in Algorithm \ref{alg1}. We divide the input into four groups equally along the channel dimension and perform HPA on the height-width, channel-height, and channel-width axes for the first three groups, respectively. For the last group, we only use DW on the feature map. Finally, we concatenate the four groups along the channel dimension and apply another DW to integrate the information from different perspectives. Note that all kernel size employed in DW are 3.


\begin{figure}[!t]
	\centerline{\includegraphics[width=12cm]{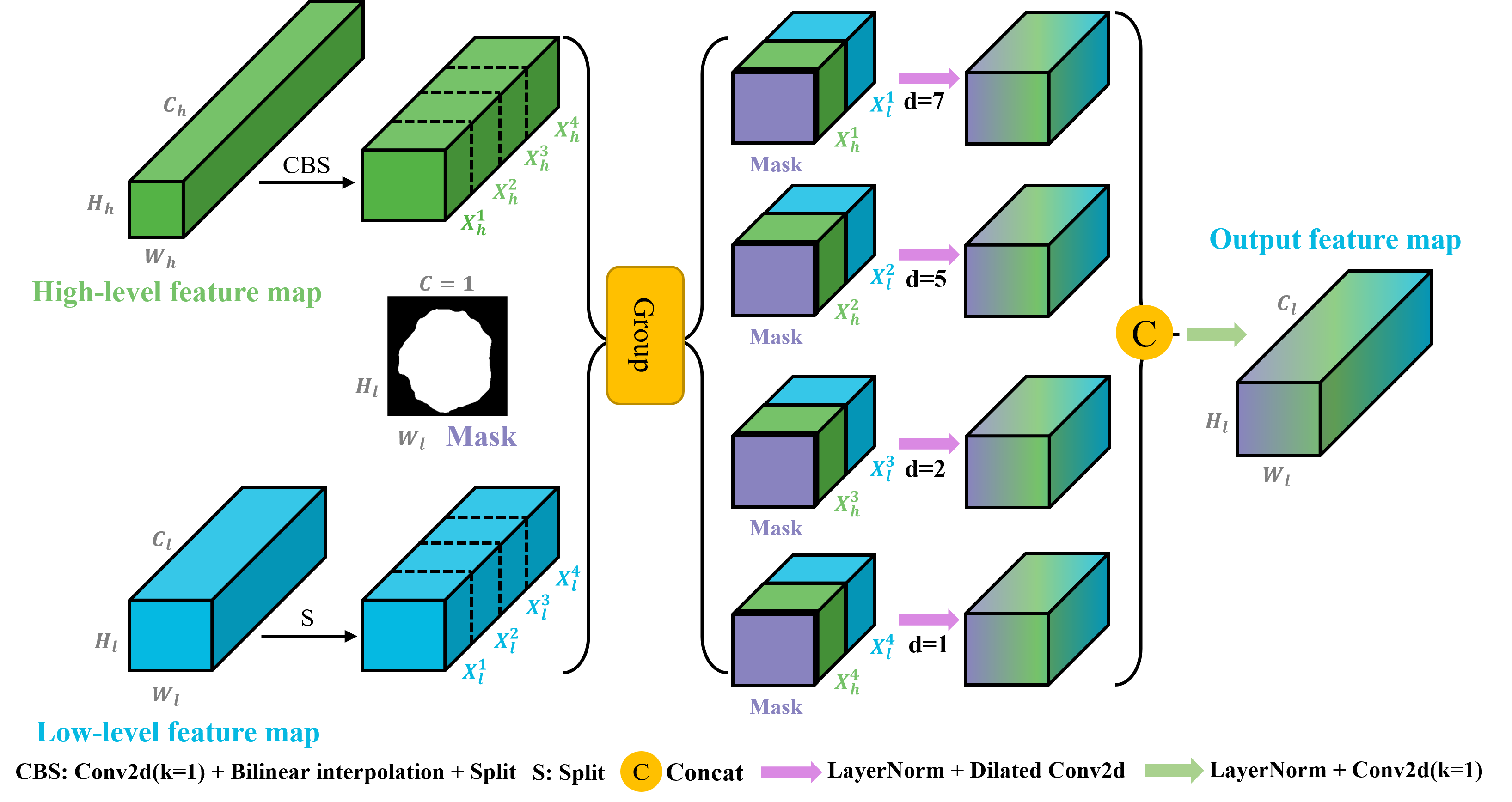}}
	\caption{The architecture of Group Aggregation Bridge module (GAB).} \label{fig3}
\end{figure}

\subsubsection{Group Aggregation Bridge module.}

The acquisition of multi-scale information is deemed pivotal for dense prediction tasks, such as medical image segmentation. Hence, as shown in Figure \ref{fig3}, we introduce GAB, which takes three inputs: low-level features, high-level features, and a mask. Firstly, depthwise separable convolution (DW) and bilinear interpolation are employed to adjust the size of high-level features, so as to match the size of low-level features. Secondly, we partition both feature maps into four groups along the channel dimension, and concatenate one group from the low-level features with one from the high-level features to obtain four groups of fused features. For each group of fused features, the mask is concatenated. Next, dilated convolutions \cite{dilatedconvolutions} with kernel size of 3 and different dilated rates of \{1, 2, 5, 7\} are applied to the different groups, in order to extract information at different scales. Finally, the four groups are concatenated along the channel dimension, followed by the application of a plain convolution with the kernel size of 1 to enable interaction among features at different scales.

\subsubsection{Loss function.}

In this study, since different GAB require different scales of mask information, deep supervision \cite{unet++} is employed to calculate the loss function for different stages, in order to generate more accurate mask information. Our loss function can be expressed as equation (1) and (2).

\begin{equation}
	l_i = Bce(y,\hat{y}) + Dice(y,\hat{y})
\end{equation}
\begin{equation}
	\mathcal{L} = \sum_{i=0}^{5}\lambda_i \times l_i
\end{equation}where $Bce$ and $Dice$ represent binary cross entropy and dice loss. $\lambda_i$ is the weight for different stage. In this paper, we set $\lambda_i$ to 1, 0.5, 0.4, 0.3, 0.2, 0.1 from $i=0$ to $i=5$ by default.

\section{Experiments}

\subsubsection{Datasets and Implementation details.} 

To assess the efficacy of our model, we select two public skin lesion segmentation datasets, namely ISIC2017 \cite{isic2017url}\cite{isic17} and ISIC2018 \cite{isic2018url}\cite{isic18}, containing 2150 and 2694 dermoscopy images, respectively. Consistent with prior research \cite{malunet}, we randomly partition the datasets into training and testing sets at a 7:3 ratio.

EGE-UNet is developed by Pytorch \cite{pytorch} framework. All experiments are performed on a single NVIDIA RTX A6000 GPU. The images are normalized and resized to 256×256. We apply various data augmentation, including horizontal flipping, vertical flipping, and random rotation. AdamW \cite{adamw} is utilized as the optimizer, initialized with a learning rate of 0.001 and the CosineAnnealingLR \cite{cosineannealingLR} is employed as the scheduler with a maximum number of iterations of 50 and a minimum learning rate of 1e-5. A total of 300 epochs are trained with a batch size of 8. To evaluate our method, we employ Mean Intersection over Union (mIoU), Dice similarity score (DSC) as metrics, and we conduct 5 times and report the mean and standard deviation of the results for each dataset.

\subsubsection{Comparative results.}

\begin{table}[!t]
	\setlength\tabcolsep{1.5pt}
	\renewcommand\arraystretch{1.25}
	\scriptsize
	\caption{Comparative experimental results on the ISIC2017 and ISIC2018 dataset.}
	\begin{center}
		\begin{tabular}{c|c|cc|cc}
			\hline
			\textbf{Dataset} &\textbf{Model}  &\textbf{Params(M)$\downarrow$}   &\textbf{GFLOPs$\downarrow$}     & \textbf{mIoU(\%)$\uparrow$}  & \textbf{DSC(\%)$\uparrow$}      \\ \hline
			\multirow{6}{*}{ISIC2017} 
			&UNet \cite{unet}    & 7.77 &13.76               & 76.98          & 86.99               \\
			&UTNetV2 \cite{utnetv2}    & 12.80 &15.50             & 77.35          & 87.23           \\
			&TransFuse \cite{transfuse}  & 26.16 &11.50            & 79.21          & 88.40      \\
			&MobileViTv2 \cite{mobilevitv2}  & 1.87 &0.70            & 78.72          & 88.09      \\
			&MobileNetv3 \cite{mobilenetv3}  & 1.19 &0.10           & 77.69         & 87.44      \\
			&UNeXt-S \cite{unext}  & 0.32 &0.10         & 78.26          & 87.80    \\
			&MALUNet \cite{malunet}  & 0.177 &0.085         & 78.78         & 88.13     \\
			&\textbf{EGE-UNet (Ours)} & \textbf{0.053} &\textbf{0.072}         & \textbf{79.81$\pm$0.10}      & \textbf{88.77$\pm$0.06}    \\ \hline
			\multirow{9}{*}{ISIC2018}
			&UNet \cite{unet}      &7.77  & 13.76             & 77.86          & 87.55                    \\
			&UNet++ \cite{unet++}   & 9.16       & 34.86     & 78.31          & 87.83                        \\
			&Att-UNet \cite{attentionunet}   &  8.73    & 16.71     & 78.43          & 87.91                      \\
			&UTNetV2 \cite{utnetv2}   &  12.80     &  15.50    & 78.97          & 88.25                        \\
			&SANet \cite{sanet}      &  23.90     &  5.96   & 79.52          & 88.59                        \\
			&TransFuse \cite{transfuse}    &  26.16   & 11.50    & 80.63          & 89.27              \\
			&MobileViTv2 \cite{mobilevitv2}  & 1.87 &0.70            & 79.88          & 88.81      \\
			&MobileNetv3 \cite{mobilenetv3}  & 1.19 &0.10           & 78.55         & 87.98      \\
			&UNeXt-S \cite{unext}    &  0.32   & 0.10   & 79.09          & 88.33              \\
			&MALUNet \cite{malunet}    &  0.177   & 0.085    & 80.25         & 89.04             \\
			&\textbf{EGE-UNet (Ours)}  &  \textbf{0.053}   & \textbf{0.072}    & \textbf{80.94$\pm$0.11}      & \textbf{89.46$\pm$0.07}    \\ \hline
		\end{tabular}
		\label{tab1}
	\end{center}
\end{table}

\begin{figure}[!t]
	\centerline{\includegraphics[width=8cm]{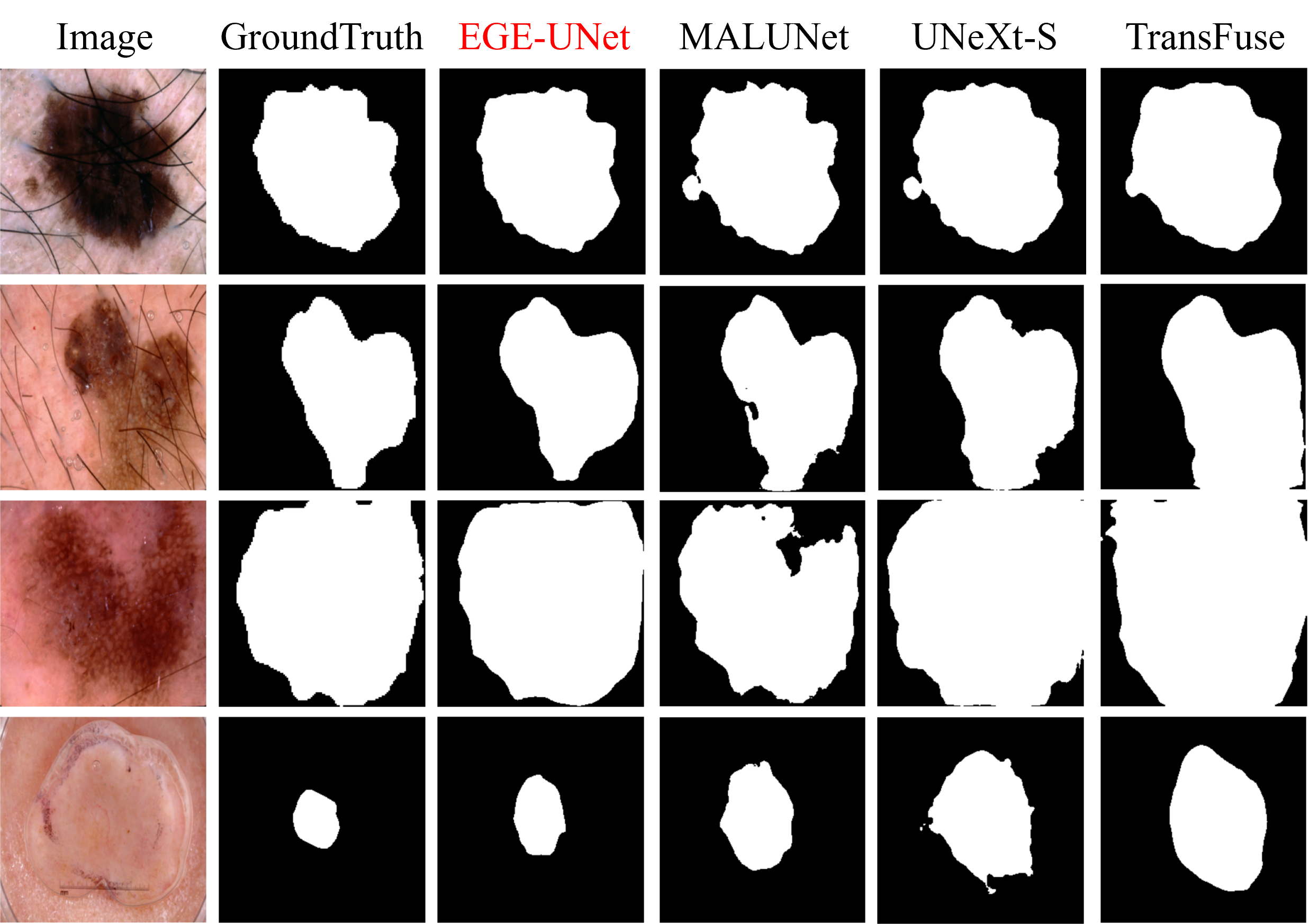}}
	\caption{Qualitative comparisons on the ISIC2018 dataset.} \label{fig4}
\end{figure}

The comparative experimental results presented in Table \ref{tab1} reveal that our EGE-UNet exhibits a comprehensive state-of-the-art performance on the \textbf{ISIC2017} dataset. Specifically, in contrast to larger models, such as TransFuse, our model not only demonstrates superior performance, but also significantly curtails the number of parameter and computation by 494x and 160x, respectively. In comparison to other lightweight models, EGE-UNet surpasses UNeXt-S with a mIoU improvement of 1.55\% and a DSC improvement of 0.97\%, while exhibiting parameter and computation reductions of 17\% and 72\% of UNeXt-S. Furthermore, EGE-UNet outperforms MALUNet with a mIoU improvement of 1.03\% and a DSC improvement of 0.64\%, while reducing parameter and computation to 30\% and 85\% of MALUNet. For the \textbf{ISIC2018} dataset, the performance of our model also outperforms that of the best-performing model. Besides, it is noteworthy that EGE-UNet is the first lightweight model reducing parameter to about 50KB with excellent segmentation performance. Figure \ref{fig1} presents a more clear visualization of the experimental findings and Figure \ref{fig4} shows some segmentation results.

\subsubsection{Ablation results.}

\begin{table}[!t]
	\setlength\tabcolsep{1.5pt}
	\renewcommand\arraystretch{1.25}
	\scriptsize
	\caption{Ablation studies on the ISIC2017 dataset. (a) the macro ablation on two modules. (b) the micro ablation on GHPA. (c) the micro ablation on GAB.}
	\begin{center}
		\begin{tabular}{c|c|cc|cc}
			\hline
			\textbf{Type} &\textbf{Model}  &\textbf{Params(M)$\downarrow$}   &\textbf{GFLOPs$\downarrow$}     & \textbf{mIoU(\%)$\uparrow$}  & \textbf{DSC(\%)$\uparrow$}      \\ \hline
			\multirow{3}{*}{(a)} 
			&Baseline & 0.107 &0.076  & 76.30 & 86.56          \\
			&Baseline + GHPA & 0.034 & 0.058  & 78.82 & 88.16          \\
			&Baseline + GAB & 0.126 & 0.086  &78.78 & 88.13     \\
			\hline
			\multirow{2}{*}{(b)} 
			&w/o multi-axis grouping & 0.074 & 0.074  & 79.13 &  88.35            \\
			&w/o DW for initialized tensor  & 0.050 & 0.072  & 79.03  & 88.29    \\
			\hline
			\multirow{2}{*}{(c)} 
			&w/o mask information & 0.052 & 0.070  & 78.97 & 88.25               \\
			&w/o dilation rate of Conv2d & 0.053 & 0.072 & 79.11 & 88.34          \\
			\hline
		\end{tabular}
		\label{tab2}
	\end{center}
\end{table}

We conduct extensive ablation experiments to demonstrate the effectiveness of our proposed modules. The baseline utilized in our work is referenced from MALUNet \cite{malunet}, which employs a six-stage U-shaped architecture with symmetric encoder and decoder components. Each stage includes a plain convolution operation with a kernel size of 3, and the number of channels at each stage is set to \{8, 16, 24, 32, 48, 64\}. In Table \ref{tab2}(a), we conduct macro ablations on GHPA and GAB. Firstly, we replace the plain convolutions in the last three layers of baseline with GHPA. Due to the efficient multi-perspective feature acquisition of GHPA, it not only outperforms the baseline, but also greatly reduces the parameter and computation. Secondly, we substitute the skip-connection operation in baseline with GAB, resulting in further improved performance. Table \ref{tab2}(b) presents the ablations for GHPA. We replace the multi-axis grouping with single-branch and initialize the learnable tensors with only random values. It is evident that the removal of these two key designs leads to a marked drop. Table \ref{tab2}(c) illustrates the ablations for GAB. Initially, we omit the mask information, and mIoU metric even drops below 79\%, thereby confirming once again the critical role of mask information in guiding feature fusion. Furthermore, we substitute the dilated convolutions in GAB with plain convolutions, which also leads to a reduction in performance.

\section{Conclusions and Future Works}

In this paper, we propose two advanced modules. Our GHPA uses a novel HPA mechanism to simplify the quadratic complexity of the self-attention to linear complexity. It also leverages grouping to fully capture information from different perspectives. Our GAB fuses low-level and high-level features and introduces a mask to integrate multi-scale information. Based on these modules, we propose EGE-UNet for skin lesion segmentation tasks. Experimental results demonstrate the effectiveness of our approach in achieving state-of-the-art performance with significantly lower resource requirements. We hope that our work can inspire further research on lightweight models for the medical image community.

Regarding limitations and future works, on the one hand, we mainly focus on how to greatly reduce the parameter and computation complexity while improving performance in this paper. Thus, we plan to deploy EGE-UNet in a real-world environment in the future work. On the other hand, EGE-UNet is currently designed only for the skin lesion segmentation task. Therefore, we will extend our lightweight design to other tasks.
\bibliographystyle{splncs04}
\bibliography{paper1980}

\end{document}